\documentclass[aps,prd,twocolumn,showpacs,superscriptaddress,amsmath,graphicx]{revtex4}

\usepackage{graphicx}

\def\slashed#1{\protect{\slash\hspace{-5pt}#1}}

\begin{document}

\title{Charm meson resonances in $D \to P \ell \nu$ decays}

\author{Svjetlana Fajfer}
\email[Electronic Address:]{svjetlana.fajfer@ijs.si}
\affiliation{J. Stefan Institute, Jamova 39, P. O. Box 3000, 1001 Ljubljana, Slovenia}
\affiliation{Department of Physics, University of Ljubljana, Jadranska 19, 
1000 Ljubljana, Slovenia}

\author{Jernej Kamenik}
\email[Electronic Address:]{jernej.kamenik@ijs.si}
\affiliation{J. Stefan Institute, Jamova 39, P. O. Box 3000, 1001 Ljubljana, Slovenia}

\date{\today}

\begin{abstract}
Motivated by recent experimental results we reconsider semileptonic $D
\to P \ell \nu_{\ell} $ decays within a model which combines heavy quark
symmetry and properties of the chiral Lagrangian.  We include excited
charm meson states, some of them recently observed, in our Lagrangian and determine their
impact on the charm meson semileptonic form factors.  We find
that the inclusion of excited charm meson states in the
model leads to a rather good agreement with the experimental results
on the $q^2$ shape of the $F_+(q^2)$ form factor. We also
calculate branching ratios for all $D \to P \ell \nu_{\ell} $ decays.
\end{abstract}

\pacs{13.20.Fc,12.39.Hg,12.39.Fe,14.40.Lb}

\maketitle

\section{Introduction}

The knowledge of the form factors which describe the weak $heavy \to
light$ semileptonic transitions is very important for the accurate
determination of the CKM parameters from the experimentally measured
exclusive decay rates. A lack of precise information about the shapes
of various form factors is thus still the main source of uncertainties
especially in these processes.
\par 
In addition to many studies of exclusive $B$-meson semileptonic
decays there are new interesting results on $D$-meson
semileptonic decays.  The CLEO and FOCUS collaborations have studied semileptonic
decays $D^0\rightarrow \pi^- \ell^+ \nu$ and $D^0\rightarrow K^- \ell^+ \nu$~\cite{Huang:2004fr,Link:2004dh}. Their data provides new information on the $D^0\rightarrow \pi^- \ell^+ \nu$ and $D^0\rightarrow K^- \ell^+ \nu$ form factors. Usually in $D$ semileptonic decays a simple pole parametrization has been used in the past. The results of Refs.~\cite{Huang:2004fr,Link:2004dh} for the single pole parameters required by the fit of their data, however, suggest pole masses, which are inconsistent with the physical masses of the lowest lying charm meson resonances. In their anlyses they also utilized a modified pole fit as suggested in~\cite{Becirevic:1999kt} and their results indeed suggest the existence of
contributions beyond the lowest lying charm meson resonances~\cite{Huang:2004fr}.
\par
There exist many theoretical calculations describing semileptonic
decays of heavy to light meson: quark models
(QM)~\cite{Wirbel:1985ji,Isgur:1988gb,Scora:1995ty,Choi:1999nu,Melikhov:2000yu,Wang:2002zb},
QCD sum rules
(SR)~\cite{Ball:1991bs,Ball:1993tp,Yang:1997ar,Khodjamirian:2000ds,Du:2003ja,Aliev:2004qd},
lattice QCD~\cite{Flynn:1997ca,Aubin:2004ej} and a few attempts to use
combined heavy meson and chiral Lagrangian theory
(HM$\chi$T)~\cite{Wise:1992hn, Casalbuoni:1996pg}. Most of the above
methods have limited range of applicability. For example, QCD sum
rules are suitable only for describing the low $q^2$ region while
lattice QCD and HM$\chi$T give good results only for the high $q^2$
region.  However, the quark models, which do provide the full $q^2$
range of the form factors, cannot easily be related to the QCD
Lagrangian and require input parameters, which may not be of
fundamental significance~\cite{Melikhov:2000yu}.
\par
Recently new experimental studies of charm meson resonances have
provided a lot of new information on the charm sector. Several
experiments on open charm hadrons have reported very interesting
discoveries: First BaBar~\cite{Aubert:2003fg} announced a new, narrow
meson $D_{sJ}(2317)^+$. This was confirmed by
Focus~\cite{Vaandering:2004ix} and CLEO~\cite{Besson:2003jp} who also
observed another narrow state, $D_{sJ}(2463)^+$. Both states were
confirmed by Belle~\cite{Krokovny:2003zq} who also provided first
evidence for two new, broad states $D_{0}^{*}(2308)$ and $D_1'(2427)$,
both ca. $350-400\mathrm{~MeV}$ higher above the usual $D^0$, $D^*$
states and with opposite parity.  Finally,
Selex~\cite{Evdokimov:2004iy} has announced a new, surprisingly narrow
state $D_{sJ}^ +(2632)$ with the spin parity assignment $1^-$. Both
$D_{sJ}(2317)^+$ and $D_{sJ}(2463)^+$ have already been proposed as
members of the $(0^+,1^+)$ spin doublet chiral partners of the
heavy-light pseudoscalar and vector $D_s$
mesons~\cite{Bardeen:2003kt,Nowak:2004uv}, while the states
$D_{0}^*(2308)$ and $D_1'(2427)$ have also been proposed as chiral
partners of the $D$-meson doublet~\cite{Nowak:2004uv}. On the other
hand, a proposal has been put forward for the $D_{sJ}^+(2632)$ state
as the first radial excitation of the
$D_s^*(2112)$~\cite{vanBeveren:2004ve,Dai:2004ng}. While strong and
electromagnetic transitions of these new states have already been
studied~\cite{Bardeen:2003kt}, they have not yet been applied to the
description of weak decays of charmed mesons. Specifically in the
description of heavy to light weak transition form factors they
are yet to be taken into account as possible transition resonances.
\par
The purpose of this paper is to investigate contributions of the newly
discovered and theoretically predicted charm mesons within an effective model based on HM$\chi$T
by incorporating the newly discovered heavy meson fields into the
HM$\chi$T Lagrangian. In this paper we only focus onto the $D
\to P \ell \nu_{\ell} $ transitions, since available experimental data can be used in
determination of all the required parameters. We restrain our
discussion to the leading chiral and $1/m_H$ terms in the expansion,
but we hope to capture the main physical features about the impact of
the nearest poles in the $t$-channel to the $q^2$-dependence of the
form factors.
\par
Within the context of heavy to light transitions the HM$\chi$T is only
valid at small recoil momentum (large $q^2$) and in order to predict
the whole $q^2$ region dependence, the form factors have to be modeled
with a presumed pole ansatz. After including excited charm vector
meson states we notice that their presence implies a natural
appearance of the double pole shape for the
$F_+$ form factor. Assuming such form factor shape we obtain good agreement with the observed $q^2$ dependence of $F_+$ in $D^0\to K/\pi$ transitions as well as with the measured branching ratios for these decays.
\par
We compare our results for the form factor $q^2$ dependence with
existing experimental~\cite{Huang:2004fr,Ablikim:2004ej,Link:2004dh} and lattice
QCD~\cite{Aubin:2004ej} results, as well as results of other
theoretical studies~\cite{Melikhov:2000yu,Wang:2002zb}. We complete
our study by calculating branching ratios for all $D
\to P \ell \nu_{\ell} $ ($P= \pi$, $K$, $\eta$, $\eta'$) transitions.
\par
Sec. II describes the framework we use in our calculations: first we
write down the HM$\chi$T Lagrangian for heavy and light mesons and
extend it to incorporate new heavy meson fields. In Sec. III we study
$q^2$ behavior of the form factors in $D \to P \ell \nu_{\ell} $ decays. 
Finally, a short summary of the results and comparison with
experimental data is given in Sec. IV.

\section{The Framework}

\subsection{Strong interactions}

Interactions, relevant for our study, between odd parity heavy mesons and light pseudoscalar
mesons are described by the leading order interaction Lagrangian,
namely (see e.g.~\cite{Wise:1992hn,Casalbuoni:1996pg})
\begin{eqnarray}
\mathcal L_{\mathrm{int}} &=& i g \langle H_b \gamma_{\mu} \gamma_{5} \mathcal A^{\mu}_{ba} \bar
 H_a \rangle,
\label{L_odd_odd}
\end{eqnarray}
with $H=1/2(1+\slashed v )[P^*_{\mu} \gamma^{\mu} - P \gamma_5]$, the
matrix representation of the heavy meson fields, where $P^*_{\mu}$ and
$P$ are creation operators for heavy-light vector and pseudo scalar
mesons respectively. Light pseudoscalar meson fields are encoded in
$\Sigma = \mathrm{exp} (2i\mathcal{M}/f)$ where $\mathcal M$ is the
pseudogoldstone flavor matrix
\begin{equation}
\mathcal M =
   \begin{pmatrix} \frac{1}{\sqrt 2} (\eta_q + \pi^0) & \pi^+ & K^+ \\
   \pi^- & \frac{1}{\sqrt 2} (\eta_q - \pi^0) & K^0 \\ K^- & \bar K^0
   & \eta_s \end{pmatrix}.
\end{equation}
To describe $\eta$ $\eta'$ mixing we follow the work of Ref.~\cite{Feldmann:1998vh}, where $|\eta\rangle = |\eta _q\rangle \cos
\phi - | \eta_s\rangle \sin \phi$ and $|\eta'\rangle = |\eta_q\rangle
\sin \phi + | \eta_s\rangle \cos \phi$, and $\phi$ is the mixing angle
between the flavor and mass eigenstates.  The light pseudoscalar meson axial current is defined as $\mathcal A_{\mu } = 1/2 (\xi^{\dagger} \partial_{\mu} \xi - \xi \partial_{\mu}
\xi^{\dagger})$, where $\xi^2=\Sigma$ so that $\mathcal
A_{\mu} \approx (i/f) \partial_{\mu}\mathcal M$. Furthermore, $\langle \ldots \rangle$ indicate a trace over spin matrices and summation over light quark flavor indices.
\par
In order to incorporate positive parity heavy meson states into the
model, we introduce the scalar-axialvector fields doublet $G=1/2(1+\slashed v )[S^*_{\mu}
\gamma^{\mu}\gamma_{5} - S]$ representing axial-vector ($S^*_{\mu}$)
and scalar ($S$) mesons and incorporate it into the interaction
Lagrangian by adding an additional leading order interaction term
between even and odd parity fields:
\begin{eqnarray}
\mathcal L'_{\mathrm{int}} &=& i h \langle G_b \gamma_{\mu} \gamma_{5} \mathcal A^{\mu}_{ba} \bar H_a \rangle + \mathrm{h.c.}.
\label{L_even_odd}
\end{eqnarray} 
\par
Finally we include the radially excited states into our discussion by introducing another odd parity heavy meson multiplet field
$H'=1/2(1+\slashed v )[P^{'*}_{\mu} \gamma^{\mu} - P'
\gamma_5]$ containing the radial excitations of ground state
pseudoscalar and vector mesons. Such excited states were predicted
in Ref.~\cite{DiPierro:2001uu}. The strong interactions between these
fields and ground state meson fields $H$ can again be described by the
lowest order interaction Lagrangian analogous to~(\ref{L_even_odd})
\begin{eqnarray}
\tilde{\mathcal L}_{\mathrm{int}} &=& i \tilde g \langle H'_b \gamma_{\mu} \gamma_{5} \mathcal A^{\mu}_{ba} \bar H_a \rangle + \mathrm{h.c.}.
\label{L_odd_oddstar}
\end{eqnarray} 

\subsection{Weak interactions}

For the semileptonic decays the weak Lagrangian can be given by the
effective current-current Fermi interaction
\begin{equation}
\mathcal L_{\mathrm{eff}} = - \frac{G_F}{\sqrt 2} \left[ \bar \ell \gamma^{\mu} (1-\gamma^5) \nu_{\ell} \mathcal J_{\mu}  \right], 
\end{equation} 
where $G_F$ is the Fermi constant and $\mathcal J_{\mu}$ is the effective
hadronic current. In heavy to light meson decays it can be written as
$\mathcal J_{\mu} = K_a J^a_{\mu}$, where constants $K^a$ parametrize the $SU(3)$
flavor mixing, while the leading order weak current $J^a_{\mu}$ in $1/m_H$ (where $m_H$ is the
heavy meson mass) and chiral expansion can be written as
\begin{eqnarray}
J_a^{\mu} &=& \frac{1}{2} i \alpha \langle \gamma^{\mu} (1-\gamma^5)
H_b \xi^{\dagger}_{ba} \rangle
\end{eqnarray}
for the $0^-$ and $1^-$ heavy mesons, while similarly for the $0^+$
and $1^+$ states we write~\cite{Casalbuoni:1996pg}
\begin{eqnarray}
J_a^{'\mu} &=& \frac{1}{2} i \alpha' \langle \gamma^{\mu} (1-\gamma^5)
G_b \xi^{\dagger}_{ba} \rangle.
\end{eqnarray}
Finally for the radially excited pseudoscalar and vector fields $J_a$
can be written as:
\begin{eqnarray}
\tilde J_a^{\mu} &=& \frac{1}{2} i \tilde\alpha \langle \gamma^{\mu} (1-\gamma^5) H'_b \xi^{\dagger}_{ba} \rangle.
\end{eqnarray}

\section{Form Factor calculation}

Now we turn to the discussion of the form factor $q^2$ distribution.
The $H\rightarrow P$ current matrix element is usually parametrized as
\begin{eqnarray}
	&&\langle P (p_P) | (V-A)^{\mu} | H (p_H) \rangle\nonumber\\*
	&&\quad= F_+(q^2) \left( (p_H+p_P)^{\mu} -
	\frac{m_{H}^2-m_{P}^2}{q^2}q^{\mu} \right) \nonumber\\*
	&&\quad\quad+ F_0(q^2) \frac{m_{H}^2-m_{P}^2}{q^2}q^{\mu},
\label{def-ff}
\end{eqnarray}
where $(V-A)$ is the weak lefthanded quark current and $q = (p_H-p_P)$
is the exchanged momentum. Here $F_+$ denotes the vector form factor
and it is dominated by vector meson resonances, while $F_0$ denotes
the scalar form factor and is expected to be dominated by scalar meson
resonance exchange~\cite{Marshak:1969tw,Wirbel:1985ji}. In order that
these matrix elements are finite at $q^2=0$, the form factors must
also satisfy the relation
\begin{equation}
F_+(0)=F_0(0)~.
\label{PP_form_factor_relations}
\end{equation}
\par
In Ref.~\cite{Becirevic:2002sc} it was pointed out that in the limit
of a static heavy meson 
\begin{equation}
|H(v)\rangle_{\mathrm{HQET}} = \lim_{m_H\to\infty} \frac{1}{\sqrt{m_H}} | H(p_H)\rangle
\end{equation}
one can use the following decomposition:
\begin{eqnarray}
&&\langle P (p_P) | (V-A)^{\mu} | H (v) \rangle_{\mathrm{HQET}} \nonumber\\*
&&\quad = \left[ p_P^{\mu} - (v\cdot p_P) v^{\mu} \right] f_p(v\cdot p_P)\nonumber\\*
&&\quad\quad + v^{\mu} f_v(v\cdot p_P),
\label{HQET_ff}
\end{eqnarray}
where the form factors $f_{p,v}$ are functions of the variable
\begin{equation}
v\cdot p_P = \frac{m_H^2 +m_P^2-q^2}{2m_H}.
\end{equation}
The form factors $F_{+,0}$ given in (\ref{def-ff}) and the form factors  
$ f_{P,v}(v\cdot p_P)$ are related to each other by matching QCD to Heavy quark 
effective theory 
(HQET) at the scale $\mu \simeq m_c$  (see Eq. (14) of Ref.~\cite{Becirevic:2002sc}). 
As in~\cite{Becirevic:2002sc} we fix the matching constants to their tree 
level values. This approach immediately accounts for the $F_{+,0}$ behavior at 
$q^2_{\mathrm{max}}$.
At the leading order in heavy quark expansion, the two definitions are then
related near zero recoil momentum ($q^2\simeq
q^2_{\mathrm{max}}=(m_H-m_P)^2$ or equivalently $|\vec p_P|\simeq0$)
as
\begin{subequations}
\begin{eqnarray}
 F_0(q^2)|_{q^2\approx q^2_{\mathrm{max}}} &=& \frac{1}{\sqrt{m_H}}
 f_v(v\cdot p_P) \label{ff_rel1}\\* F_+(q^2)|_{q^2\approx
 q^2_{\mathrm{max}}} &=& \frac{\sqrt{m_H}}{2} f_p(v\cdot
 p_P),\label{ff_rel2}
\end{eqnarray}
\end{subequations}
\par
Note, however, that the $F_+$ and $F_0$ form factors might contain  
$1/m_c$ corrections~\cite{Becirevic:2002sc}, which we do not consider 
within present approach. 
\par
We use Feynman rules for HM$\chi$T coming from the strong and weak Lagrangians 
described in the previous section.
 For the heavy meson
propagators we use $i\delta_{ab}/2(v\cdot k-\Delta)$ and $-i
\delta_{ab} (g_{\mu\nu}-v_{\mu} v_{\nu})/2(v\cdot k - \Delta)$ for the
pseudoscalar(scalar) and vector (axial) mesons respectively, where
$k^{\mu}=q^{\mu}-m_{H} v^{\mu}$ and $\Delta = \Delta_R$ is the mass
splitting between the heavy resonance meson $R$ and the ground state
heavy pseudoscalar meson (see e.g. Ref.~\cite{Casalbuoni:1996pg}). For the hadronic current matrix element we
thus get
\begin{eqnarray}
&&\langle P(p_P) | J^{\mu} | H(v) \rangle = -\frac{\alpha}{f} v^{\mu}
\nonumber\\* &&\quad +\frac{\alpha}{f} g \frac{v^{\mu} v\cdot
p_P-p_P^{\mu}}{v\cdot p_P + \Delta_{H^*}} \nonumber\\* &&\quad +
\frac{\tilde\alpha}{f} \tilde g \frac{v^{\mu} v\cdot
p_P-p_P^{\mu}}{v\cdot p_P + \Delta_{H^{'*}}} + \frac{\alpha'}{f} h
\frac{v^{\mu} v\cdot p_P}{v \cdot p_P + \Delta_{H_S}}.
\label{J_HMCT}
\end{eqnarray} 
We apply the projectors $v_{\mu}$ and 
$v_{\mu} v\cdot p_P - p_{P\mu}$ on eq. (\ref{J_HMCT})  and extract 
 the form factors $F_{+}(q^2)$ and $F_0(q^2)$ at
$q^2_{\mathrm{max}}$ using Eqs.~(\ref{HQET_ff}), (\ref{ff_rel1})
and~(\ref{ff_rel2}):
\begin{eqnarray}
F_+(q^2_{\mathrm{max}}) &=& - \frac{\alpha}{2\sqrt m_H f} g
\frac{m_H}{m_P + \Delta_{H^*}} \nonumber\\* && -
\frac{\tilde\alpha}{2\sqrt m_H f} \tilde g \frac{m_H}{m_P +
\Delta_{H^{'*}}},\label{F+FFq}
\end{eqnarray} 
and
\begin{eqnarray}
F_0(q^2_{\mathrm{max}}) &=& -\frac{\alpha}{\sqrt m_H f} +
\frac{\alpha'}{\sqrt m_H f} h \frac{m_P}{m_P +
\Delta_{H_S}}.\label{F0FFq}
\end{eqnarray} 
\par
If one uses directly relation~(\ref{def-ff}) instead of this
extraction of form factors at large $q^2_{\mathrm{max}}$~\cite{Becirevic:2002sc} one ends up with the mixed leading $\sqrt
m_H$ terms and the subleading $1/\sqrt m_H$ terms. Furthermore, the
scalar meson contribution appears in the $F_+$ form factor.  The
extraction of form factors we follow here~\cite{Becirevic:2002sc}
gives a correct $1/m_H$ behavior of the form factors and the
contributions of $1^-$ resonances enter in $F_+$, while $0^+$
resonances contribute to $F_0$ as they must~\cite{Marshak:1969tw}.  In previous theoretical
studies of the $D$ meson form factors~\cite{Casalbuoni:1996pg} a single pole
parametrization has been used when extrapolating from $F_{+}(q^2_{\mathrm{max}})$. In such
calculations only the $D^*$ resonance contributed, and a lower value
of the $g$ strong coupling was used. At that time only few decay
rates were measured. In comparison with the present experimental data
the predicted branching ratios were too large. The approach of Ref.~\cite{Bajc:1995km} was developed to treat $D$ meson semileptonic decay within heavy light
meson symmetries in the allowed kinematical region by using the full
propagators.  We find that this approach cannot reproduce the
observed  $q^2$ shape of the $F_+$ form factors.
\par
In our study of form factors' $q^2$ distributions, we
follow the analysis of Ref.~\cite{Becirevic:1999kt}, where the $F_+$
form factor is given as a sum of two pole contributions, while the
$F_0$ form factor is written as a single pole. This 
parametrization includes all known properties of form factors at large $m_H$ and the 
QCD sum rules for low $q^2$ region. Their proposal for the form factor 
parametrization is
\begin{subequations}
\begin{eqnarray}
F_+(q^2)&=&c_B\left(\frac{1}{1-x}-\frac{a}{1-x/\gamma}\right)\label{F+FF}\\
F_0(q^2)&=&\frac{c_B(1-a)}{1-b x},\label{F0FF}
\end{eqnarray}
\end{subequations}
where $x=q^2/m_{H^*}^2$. 
 Using the relation which connects the 
form factors within large energy
release approach  ~\cite{Charles:1998dr} 
the authors in Ref.~\cite{Becirevic:1999kt}
then  find $a=1/\gamma$ and so obtain a simplified double pole
function for the $F_+$ form factor
\begin{equation}
F_+(q^2)=\frac{F_+(0)}{(1-x)(1-a x)}\label{f_+_dipole}\\
\end{equation}
Although the $D$ mesons might not be  considered heavy enough, 
 we employ these formulas with the model matching condition at
$q^2_{\mathrm{max}}$. 
At the same time we fix the
parameters $a$ and $b$ in Eqs.~(\ref{F0FF}) and~(\ref{f_+_dipole}) by the
next-to-nearest resonances. We use physical pole masses of excited state charmed mesons in this extrapolation, giving for the ratios $a=m_{H^{'*}}^2/m_{H^{*}}^2$ and
$b=m_{H_{S}}^2/m_{H^{*}}^2$.
Although in the original idea~\cite{Becirevic:1999kt} the extra pole in $F_+$ parametrized all the neglected higher resonances, we are here saturating those by a single nearest resonance.
\par
In our numerical analyses we use available experimental information and theoretical predictions on charm meson resonances. Particularly for the
scalar resonance we use the $D_{sJ}(2317)^+$ state with mass
$m_{D_{sJ}(2317)^+}=2.317~\mathrm{GeV}$. For the radially excited vector resonance we then have the Selex $D_{sJ}^+(2632)$ state with mass $m_{D_{sJ}^{+}(2632)}=2.632~\mathrm{GeV}$. It is important to note, however, that so-far the Selex discovery has not been confirmed by any other searches. In $D$ decays, the situation is similarly ambiguous. Although the vector $D^{'*}$ resonance was discovered by Delphi~\cite{Abreu:1998vk} with a mass of
$m_{D^{'*}}=2.637~\mathrm{GeV}$ and spin-parity $1^-$, its existence
was not confirmed by other searches~\cite{DiPierro:2001uu}. On the
other hand recent theoretical studies~\cite{DiPierro:2001uu,Vijande:2003uk} indicate that both
radially excited states of $D$ as well as $D_s$ should have slightly
larger masses of $m_{D^{'*}}\simeq2.7~\mathrm{GeV}$ and $m_{D_{s}^{'*}}\simeq2.8~\mathrm{GeV}$~\cite{DiPierro:2001uu}. We use these theoretically predicted values in our analysis. Even less clear is the situation with the scalar resonance ($D'$), which was predicted in Ref.~\cite{DiPierro:2001uu}  but not yet confirmed by experiment. Here we use the theoretically predicted mass value of $m_{D'}=2.3~\mathrm{GeV}$ from Ref.~\cite{DiPierro:2001uu}.
\par
In our  calculations we use for the heavy meson weak
current coupling $\alpha = f_H \sqrt{m_H}$~\cite{Becirevic:2002sc},
which we calculate from the lattice QCD value of $f_{D} = 0.225~\mathrm{GeV}$~\cite{Simone:2004fr}
and experimental $D$ meson mass $m_{D}=1.87~\mathrm{GeV}$~\cite{Eidelman:2004wy} yielding
$\alpha=0.33~\mathrm{GeV^{3/2}}$. For light pseudoscalar mesons we use
$f = 130~\mathrm{MeV}$, while for the $\eta-\eta'$ mixing angle
we use the value of $\phi\simeq40^{\circ}$~\cite{Feldmann:1998vh}. For
the $g$-coupling we use the experimentally determined value of
$g=0.59$~\cite{Anastassov:2001cw} while for the $h$ coupling we use
the value of $h=-0.6$ from the global estimate of
Ref.~\cite{Becirevic:2004uv}.  We can also derive a constraint on the
absolute value of $\tilde g$ from the Selex bound on the
$D_{sJ}(2632)\to D^0 K^+$ decay rate of $\Gamma (D_{sJ}(2632)\to D^0
K^+) < 17~\mathrm{MeV}$~\cite{Evdokimov:2004iy}. In HM$\chi$T this
decay rate can be derived directly from the interaction
Lagrangian~(\ref{L_odd_oddstar}) and reads $\Gamma = {1}/{6\pi}
[{\tilde g}/{f}]^2 |\vec p_K|^3$ which gives for the coupling constant
$|\tilde g| < 0.21$.
\par
Eqs.~(\ref{F+FF})~(\ref{F0FF}) and~(\ref{f_+_dipole}) can be combined
to obtain a theoretical estimate for the value of $\tilde g \tilde
\alpha$ in the limit of infinite heavy meson mass. By equating both
terms in Eqs.~(\ref{F+FFq}) and~(\ref{F+FF}) at $q^2_{\mathrm{max}}$
and then imposing $a=1/\gamma$ one obtains
\begin{equation}
\tilde\alpha\tilde g \sim  - 
\alpha g \frac{m_P+\Delta_{H^{'*}}}{m_P+\Delta_{H^*}}\times 
\frac{(m_H-m_P)^2-m_{H^{*}}^2}{(m_H-m_P)^2-m_{H^{'*}}^2}.\label{agtilde}
\end{equation}
We apply this formula to the $D\to\pi$ transitions where the chiral symmetry
breaking corrections are smallest and thus the results most reliable. 
This yields $\tilde\alpha\tilde g \sim -0.15~\mathrm{GeV}^{3/2}$.
On the other hand the $1/m_D$  and chiral corrections might still modify this result significantly. Such corrections were explicitely written out in Ref.~\cite{Boyd:1994pa} but they include additional parameters which cannot be fixed within this context.
\par
Similarly,  in this limit we can infer on the value of $\alpha'$. By
applying equalities~(\ref{PP_form_factor_relations})
and~(\ref{agtilde}) to Eqs.~(\ref{F0FFq}) and~(\ref{F0FF}) we obtain
\begin{widetext}
\begin{equation}
\alpha' \sim \frac{\alpha}{h} \times \frac{m_P+\Delta_{H_S}}{m_P} \left[ 1- \frac{g}{2}\times \frac{m_H}{m_P+\Delta_{H^*}} \left( 1- \frac{m_{H^*}^2}{m_{H^{'*}}^2}\right) \frac{1-\frac{(m_H-m_P)^2}{m_{H^*}^2}}{1-\frac{(m_H-m_P)^2}{m_{H_S}^2}} \right],
\label{eq_alpha_s}
\end{equation}
\end{widetext}
which gives, when applied to the $D\to\pi$ transitions a value of
$\alpha' \sim -0.69~\mathrm{GeV}^{3/2}$.
\par
Alternatively the values of the new model parameters can be
determined by fitting the model predictions to known experimental
values of branching ratios $\mathcal B (D^0\rightarrow K^- \ell^+
\nu)$, $\mathcal B (D^+\rightarrow \bar K^0 \ell^+ \nu)$, $\mathcal B
(D^0\rightarrow \pi^- \ell^+ \nu)$, $\mathcal B (D^+\rightarrow \pi^0
\ell^+ \nu)$, $\mathcal B (D^+_s\rightarrow \eta \ell^+ \nu)$ and
$\mathcal B (D^+_s\rightarrow \eta' \ell^+
\nu)$~\cite{Eidelman:2004wy}. In our decay width calculations we shall
neglect the lepton mass, so the form factor $F_0$, which is
proportional to $q^{\mu}$, does not contribute. For the decay width we
get~\cite{Bajc:1995km}:
\begin{equation}
\Gamma = \frac{G_F^2 m_H^2 |K_{HP}|^2}{24 \pi^3} \int_0^{y_m} dy | F_+(m_H^2 y) |^2 |\vec p_P (y)|^3,
\end{equation} 
where $y=q^2/m_H^2$, so that
\begin{equation}
y_m = \left(1-\frac{m_P}{m_H} \right)^2
\end{equation} 
and
\begin{equation}
|\vec p_P (y)|^2 = \frac{[m_H^2 (1-y) + m_P^2]^2}{4 m_H^2} -m_P^2.
\label{3p_P}
\end{equation} 
The constants $K_{HP}$ parametrize the flavor mixing relevant to a
particular transition, and are given in Table~\ref{PP_mixing_table}
together with the pole mesons.
\begin{table}
\caption{\label{PP_mixing_table} The pole mesons and the flavor mixing constants $K_{HP}$ for the $D\rightarrow P$ semileptonic decays.}
\begin{ruledtabular}
\begin{tabular}{cccccc}
	$H$ & $P$ & $H^*$ & $H'^*$ & $H_S$ & $K_{HP}$\\\hline 
	$D^0$ & $K^-$ & $D^{*+}_s$ & $D'^{*+}_{s}$ & $D_{sJ}(2317)^+$ & $V_{cs}$\\
	$D^+$ & $\bar K^0$ & $D^{*+}_s$ & $D'^{*+}_{s}$ & $D_{sJ}(2317)^+$ & $V_{cs}$\\ 
	$D^+_s$ & $\eta$ & $D^{*+}_s$ & $D'^{*+}_{s}$ & $D_{sJ}(2317)^+$ & $ V_{cs} \sin\phi$\\ 
	$D^+_s$ & $\eta'$ & $D^{*+}_s$ & $D'^{*+}_{s}$ & $D_{sJ}(2317)^+$ & $V_{cs} \cos\phi$\\ 
	$D^0$ & $\pi^-$ & $D^{*+}$ & $D'^{*+}$ & $D'^+$ & $V_{cd}$\\ 
	$D^+$ & $\pi^0$ & $D^{*+}$ & $D'^{*+}$ & $D'^+$ & $V_{cd}/\sqrt 2$\\ 
	$D^+$ & $\eta$ & $D^{*+}$ & $D'^{*+}$ & $D'^+$ & $V_{cd}\cos\phi/\sqrt 2$\\ 
	$D^+$ & $\eta'$ & $D^{*+}$ & $D'^{*+}$ & $D'^+$ & $V_{cd}\sin\phi/\sqrt 2$\\ 
	$D^+_s$ & $K^0$ & $D^{*+}$ & $D'^{*+}$ & $D'^+$ & $V_{cd}$\\
\end{tabular}
\end{ruledtabular}
\end{table}
\par
We calculate the result for $\tilde g \tilde \alpha$ by a weighted
average of values obtained from all the measured decay rates taking account for the experimental
uncertainties. The calculation yields $\tilde\alpha\tilde g = -0.0050\mathrm{~GeV^{3/2}}$, which is rather small compared to estimation given by (\ref{agtilde}). 
This discrepancy can be attributed  to the presence of $1/m_D$ and chiral corrections which are not included systematically into consideration here due too many new parameters~\cite{Boyd:1994pa} which cannot be fixed within this approach. However, we estimate the influence of such corrections on the fitted value of $\tilde\alpha\tilde g$ by varying the value of $\alpha g / f$ in Eq.~(\ref{F+FFq}) by $10\%$~\cite{Stewart:1998ke} and inspecting the fit results. We obtain a range of $\tilde\alpha\tilde g \in [-0.05,0.04]~\mathrm{GeV}^{3/2}$. 
\par
Knowing $\tilde\alpha\tilde g$ we can further infer on the value of $\alpha'$. The equality~(\ref{PP_form_factor_relations}), when applied to the same decays with $\tilde\alpha\tilde g = -0.0050\mathrm{~GeV^{3/2}}$, gives an average value of $\alpha' = -0.47\mathrm{~GeV^{3/2}}$, which is close to the estimate from~(\ref{eq_alpha_s}). We use these values for $\tilde\alpha\tilde g$ and $\alpha'$ in all our subsequent analyses.
\par
\begin{figure}[H]
\scalebox{0.9}{\includegraphics{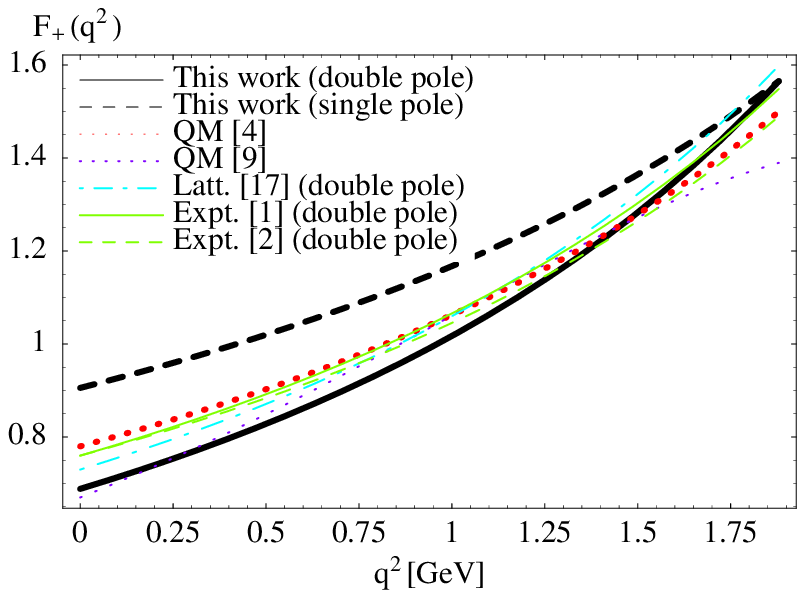}}
\caption{\label{FplotDK}Comparison of the  $D^0 \rightarrow K^-$ transition $F_+$ form factor $q^2$ dependence of our model double pole extrapolation (thick solid (black) line), single pole extrapolation (thick dashed (black) line), quark model of Ref.~\cite{Melikhov:2000yu} (thick dotted (magenta) line), quark model of Ref.~\cite{Wang:2002zb} (thin dotted (purple) line), lattice QCD fitted to a double pole~\cite{Aubin:2004ej} (dot-dashed (blue) line) and experimental double pole fits~\cite{Huang:2004fr,Link:2004dh} (thin (green) solid and dashed lines).}
\end{figure}
\begin{figure}[H]
\scalebox{0.9}{\includegraphics{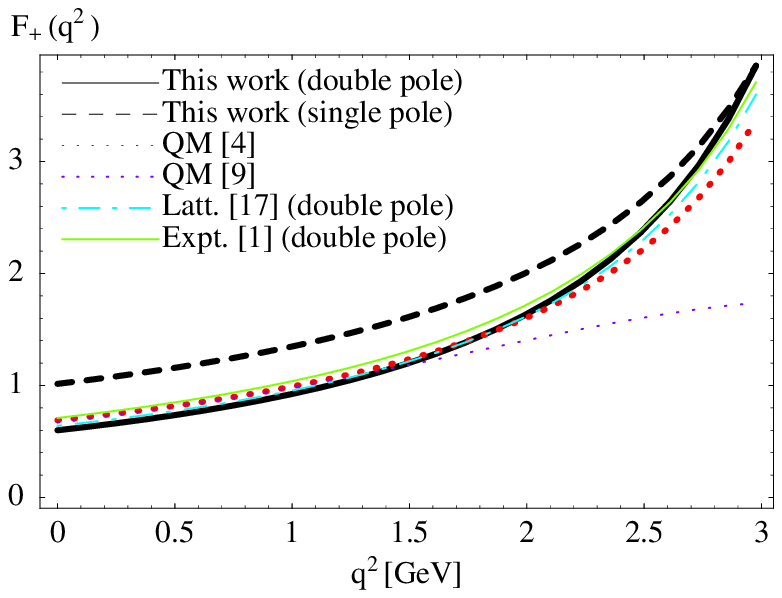}}
\caption{\label{FplotDPi}Comparison of $D^0 \rightarrow \pi^-$ transition $F_+$ form factor $q^2$ dependence of our model double pole extrapolation (thick solid (black) line), single pole extrapolation (thick dashed (black) line), quark model of Ref.~\cite{Melikhov:2000yu} (thick dotted (magenta) line), quark model of Ref.~\cite{Wang:2002zb} (thin dotted (purple) line), lattice QCD fitted to a double pole~\cite{Aubin:2004ej} (dot-dashed (blue) line) and experimental double pole fit~\cite{Huang:2004fr} (thin solid (green) line).}
\end{figure}

We next draw  the $q^2$ dependence of the  $F_+$ form factors for the $D^0\rightarrow K^-$ and $D^0\rightarrow \pi^-$ transitions  and compare it with  results of quark models in Refs.~\cite{Melikhov:2000yu,Wang:2002zb},  lattice QCD double pole fit analysis~\cite{Aubin:2004ej}, as well as  the experimental results of a double pole fit from CLEO~\cite{Huang:2004fr} and FOCUS~\cite{Link:2004dh} with $F_+(0)$ values taken from Ref.~\cite{Ablikim:2004ej}. The results are depicted in FIG.~\ref{FplotDK} and FIG.~\ref{FplotDPi}.
For comparison we also plot results when single pole fit is used. Also
in this case we calculate $F_{+} (q^2_{\mathrm{max}})$ within
HM$\chi$T and take into account both resonances
(Eq.~(\ref{F+FFq})). From the plots it becomes apparent, that our
model's predictions for both $D^0\to K^-$ and $D^0\to \pi^-$
transition $F_+$ form factors are in good agreement with
experimental and lattice data when extrapolated with a double pole,
while single pole extrapolations are not in good agreement with
experimental results. This discrepancy further increases if only the
first resonance contribution is kept in the $F_{+}
(q^2_{\mathrm{max}})$ calculation for the single pole extrapolation
within HM$\chi$T as done in previous
calculations~\cite{Casalbuoni:1996pg}. Note also that the experimental fits on the single pole parametrization of the $F_+$ form factor in $D^0\to\pi^- (D^0\to K^-)$ transitions done in Refs.~\cite{Huang:2004fr,Link:2004dh} yielded effective pole masses which are somewhat lower than the physical masses of the $D^* (D^{*}_s)$ meson resonances used in this analysis.
\par
\begin{figure}[H]
\scalebox{0.9}{\includegraphics{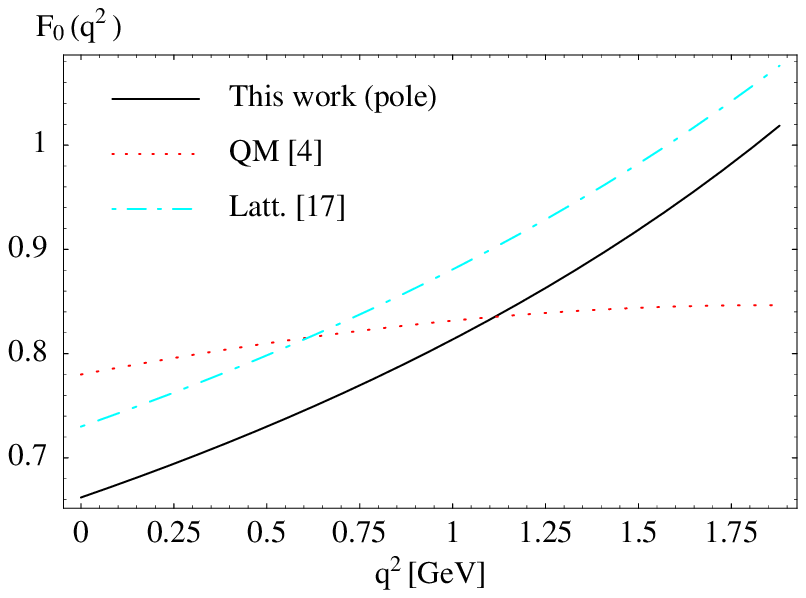}}
\caption{\label{F0plotDK}Comparison of the  $D^0 \rightarrow K^-$ transition $F_0$ form factor $q^2$ dependence of our model (solid (black) line), 
quark model of Ref.~\cite{Melikhov:2000yu} (dotted (magenta) line) and lattice QCD fitted to a pole~\cite{Aubin:2004ej} (dot-dashed (blue) line).}
\end{figure}
\begin{figure}[H]
\scalebox{0.9}{\includegraphics{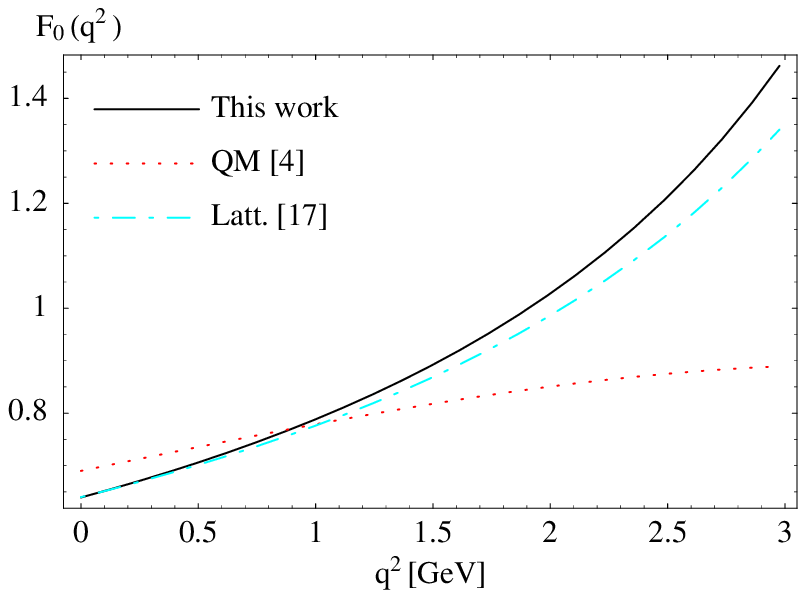}}
\caption{\label{F0plotDPi}Comparison of the  $D^0 \rightarrow \pi^-$ transition $F_0$ form factor $q^2$ dependence of our model (solid (black) line), 
quark model of Ref.~\cite{Melikhov:2000yu} (dotted (magenta) line) and lattice QCD fitted to a pole~\cite{Aubin:2004ej} (dot-dashed (blue) line).}
\end{figure}

We also compare our predictions for the $F_0$ scalar  form factor
$q^2$ dependence for the $D^0\rightarrow K^-$ and $D^0\rightarrow
\pi^-$ transitions with those of quark model in
Ref.~\cite{Melikhov:2000yu} and with lattice QCD pole fit
analysis of Ref.~\cite{Aubin:2004ej}. The results are depicted in
FIG.~\ref{F0plotDK} and FIG.~\ref{F0plotDPi}.
Note that without the scalar resonance, one only gets a contribution from the $\alpha / \sqrt m_H f$ term from Eq.~(\ref{F0FFq}). This gives for the $q^2$ dependence of $F_0$ a constant value $F_0(q^2)=1.81$ for both $D^0\to\pi^-$ and $D^0\to K^-$ transitions, which largely disagrees with lattice QCD results as well as heavily violates relation~(\ref{PP_form_factor_relations}).
\par
Finally, using numerical values as explained above, we calculate the
branching ratios for all the relevant $D\rightarrow P$ semileptonic
decays and compare the predictions of our model with various other
model predictions found in the literature, and with experimental data
from PDG. The results are summarized in Table~\ref{PP_results_table}.
\begingroup
\squeezetable
\begin{table}[H]
\caption{\label{PP_results_table} The branching ratios for the $D\rightarrow P$ semileptonic decays. Comparison of different model predictions with experiment as explained in the text.}
\begin{ruledtabular}
\begin{tabular}{ccc}
   Decay & $\mathcal{B}[\%]$ & Reference \\\hline 
   
   $D^0\rightarrow K^-$
   & $3.4$ & This work (double pole) \\ & $4.9$ & This work (single
   pole) \\ & $3.75\pm1.16$ & QM~\cite{Choi:1999nu} \\ & $4.0$ &
   QM~\cite{Melikhov:2000yu} \\ & $3.9\pm1.2$ & QM~\cite{Wang:2002zb}
   \\ & $2.7\pm0.6$ & SR~\cite{Ball:1991bs} \\ & $3.4{+1.2 \atop
   -1.0}$ & SR~\cite{Yang:1997ar} \\ & $3.7 \pm 1.4$ & SR~\cite{Khodjamirian:2000ds}
   \\ & $3.43\pm0.14$ & Expt.\\
   \\[-5pt] 
   
   $D^0\rightarrow \pi^-$ & $0.27$ & This work (double pole)
   \\ & $0.56$ & This work (single pole) \\ & $0.236\pm0.034$ &
   QM~\cite{Choi:1999nu} \\ & $0.39$ & QM~\cite{Melikhov:2000yu} \\ &
   $0.30\pm0.09$ & QM~\cite{Wang:2002zb} \\ & $0.16\pm0.3$ &
   SR~\cite{Ball:1993tp} \\ & $0.28{+0.09 \atop -0.08}$ &
   SR~\cite{Yang:1997ar} \\ & $0.27 \pm 0.10$ & SR~\cite{Khodjamirian:2000ds}
   \\ & $0.36\pm0.06$ & Expt. \\ \\[-5pt] 
   
   $D^+_s\rightarrow \eta$ &
   $1.7$ & This work (double pole) \\ & $2.5$ & This work (single
   pole) \\ & $1.8\pm0.6$ & QM~\cite{Choi:1999nu} \\ & $2.45$ &
   QM~\cite{Melikhov:2000yu} \\ & $2.5\pm0.7$ & Expt. \\ \\[-5pt]
   
   $D^+_s\rightarrow \eta'$ & $0.61$ & This work (double pole) \\ &
   $0.74$ & This work (single pole) \\ & $0.93\pm0.29$ &
   QM~\cite{Choi:1999nu} \\ & $0.95$ & QM~\cite{Melikhov:2000yu} \\ &
   $0.89\pm0.33$ & Expt. \\ \\[-5pt] 
   
   $D^+\rightarrow \bar K^0$ & $8.4$
   & This work (double pole) \\ & $12.4$ & This work (single pole) \\
   & $6.8\pm0.8$ & Expt. \\ \\[-5pt] 
   
   $D^+\rightarrow \pi^0$ & $0.33$ &
   This work (double pole) \\ & $0.70$ & This work (single pole) \\ &
   $0.31\pm0.15$ & Expt. \\ \\[-5pt] 
   
   $D^+\rightarrow \eta$ & $0.10$ &
   This work (double pole) \\ & $0.15$ & This work (single pole) \\ &
   $<0.5$ & Expt. \\ \\[-5pt] 
   
   $D^+\rightarrow \eta'$ & $0.016$ & This
   work (double pole) \\ & $0.019$ & This work (single pole) \\ &
   $<1.1$ & Expt. \\ \\[-5pt] 
   
   $D^+_s\rightarrow K^0$ & $0.20$ & This
   work (double pole) \\ & $0.32$ & This work (single pole) \\ & $0.3$
   & QM~\cite{Melikhov:2000yu} \\
\end{tabular}
\end{ruledtabular}
\end{table}
\endgroup
For comparison we also include the results for the rates obtained with
our approach for $F_+(q_{\mathrm{max}}^2)$ (Eq.~\ref{F+FFq}) but using a single pole
fit. It is very interesting that our model extrapolated with a double
pole gives branching ratios for $D \to P \ell \nu_{\ell}$ in rather good
agreement with experimental results for the already measured decay
rates, while the predictions for the unmeasured decay rates agree with results of existing approaches. It is also obvious that the single pole fit gives the rates up to a factor of two larger than the experimental results. Only for decays to $\eta$ and $\eta'$ as given in  Table~\ref{PP_results_table}, an agreement with experiment of the double pole version of the model is not better but worse than for the single pole case.
\par
We also calculate branching ratios within our model, when the value of $\tilde\alpha\tilde g$ is estimated from Eq.~(\ref{agtilde}) rather then from the branching ratio fit. While the $D\to\pi$ decay rates remain in reasonable agreement with experimental values, predictions for other decay rates come out up to 30\% lower. This again indicates large contributions from $1/m_D$ and especially chiral corrections in these decay amplitudes. We also attempt using experimental fits for the parameter $a$ in Eq.~(\ref{f_+_dipole}) instead of saturating the second pole by physical pole masses $a=m_{H^*}^2/m_{H'^*}^2$. By using the experimentally fitted values of $a=0.36(0.37)$ for $D\to K(D\to\pi)$ decays from Ref.~\cite{Huang:2004fr} we notice that, while the fitted values of $\tilde\alpha\tilde g$ tend to shift towards larger negative values, the overall goodness of fit on the experimental branching ratios and consequently decay rate predictions of our model remain almost the same. This indicates that the actual positions of the poles (masses of excited charmed resonances) are not very volatile input parameters of our model. On the other hand corrections of the order $1/m_D$, as well as chiral corrections in the case of $D \to K \ell \nu_{\ell}$ and $D_s \to \eta (\eta') \ell \nu_{\ell}$ might improve the agreement with the experimental data, but due to the presence of a large number of new couplings it is impossible to include them into the calculation within present framework. We expect that the errors in the predicted decays rates stemming from the uncertainties in the input parameters we used can be $~ 30\%$.

\section{Summary}

We have investigated semileptonic form factors for $D \to P$ decays
within an approach which combines heavy meson and chiral symmetry. The
contributions of excited charm meson states are
included into analysis. The double pole behavior of the $F_+$ form factors
is a result of the presence of the two $1^-$ charm meson resonances. 
The obtained $q^2$ dependence of the form factors is in good agreement with
recent experimental results and the lattice calculation for $D^0 \to K^-
(\pi^-) \ell \nu_{\ell}$. The calculated branching ratios are close to the experimental ones. 
\par
Our study shows that the single physical pole parametrization cannot explain the
$q^2$ distribution of the semileptonic form factors. 
The calculated rates for this form factor parametrization are not in 
good agreement with the experimental results.

\begin{acknowledgments}
We thank D. Be\'cirevi\'c for many very fruitful discussions and his
comments on the manuscript. We would also like to thank D. Y. Kim and A. Khodjamirian for their comments. This work is supported in part by the
Ministry of Education, Science and Sport of the Republic of Slovenia.
\end{acknowledgments}

\bibliography{article}

\end{document}